\documentclass[conference]{IEEEtran}
\usepackage{cite}
\usepackage{psfrag}

\ifCLASSINFOpdf
\usepackage[pdftex]{graphicx}
\graphicspath{{./img/}}
\DeclareGraphicsExtensions{.pdf,.jpeg,.png,.eps}
\else
\usepackage[dvips]{graphicx}
\graphicspath{{./img/}}
\DeclareGraphicsExtensions{.eps}
\fi

\usepackage{balance}
\usepackage[cmex10]{amsmath}
\usepackage{algorithmic}

\usepackage{array}

\usepackage{mdwmath}
\usepackage{mdwtab}

\usepackage[caption=false,font=footnotesize]{subfig}
\usepackage{url}

\begin{document}
\title{Equilibrium Properties of Rate Control Protocols}

\author{\IEEEauthorblockN{Abhijit Kiran Valluri}
\IEEEauthorblockA{University of Maryland, College Park, MD, USA\\
Email: abhijitv@umd.edu}}
\maketitle

\begin{abstract}
We analyze the stability of the Rate Control Protocol (RCP) using two different models that have been
proposed in literature. Our objective is to better understand the impact of the protocol parameters 
and the effect different forms of feedback have on the stability of the network.
We also highlight that different time scales, depending on the propagation delay relative to the queuing delay, 
have an impact on the nonlinear and the stochastic properties of the protocol fluid models. To better understand 
some of the nonlinear properties, we resort to local bifurcation analysis where we exhibit the existence of a
Hopf type bifurcation that then leads to stable limit cycles. Our work serves as a step towards a more comprehensive 
understanding of the nonlinear fluid models that have been used as representative models for RCP. 
\end{abstract}
\IEEEpeerreviewmaketitle

\section{Introduction}

Congestion in the Internet is just as critical now as ever before, with the proliferation of
high speed mobile and broadband internet and the large increase in demand for several high bandwidth 
applications such as multimedia streaming and real-time communications. Understandably, the Internet
community is actively working to develop congestion control for such applications, through the 
``RTP Media Congestion Avoidance Techniques'' (RM-CAT) working group\footnote{URL: https://tools.ietf.org/wg/rmcat/}.

Real-time applications perform poorly in the presence of TCP cross-traffic, due to the inherent fairness 
issues that arise from TCP's Additive Increase Multiplicative 
Decrease (AIMD) congestion control algorithms. Numerous studies have exhibited that standard TCP AIMD protocols 
\cite{tcp1, tcp2} are unsuitable for next generation networks, and repeated attempts have been made to improve it.
Explicit Congestion control Protocols (ECPs) such as \cite{xcp} aim to provide better fairness. 
The need for more explicit feedback is well recognized which in turn motivates the requirement for a comprehensive 
theoretical framework within which to design transport protocols. Therefore, in this paper, we reconsider the Rate 
Control Protocol (RCP) \cite{NanditaPhD} that has been proposed earlier and study its stability properties.

RCP has received significant attention from the research community: \cite{vr} develops some stability properties 
of a  max-min RCP in small buffer regime, \cite{nanswitch} computationally develops some sufficient 
conditions for local stability, \cite{bufferrcp} considers a dynamic environment with RCP flows arriving 
and departing over a single link, \cite{krv} develops an $\alpha$-fair variant of RCP and investigates 
some of the associated local stability properties, \cite{netfpga} develops a NetFPGA hardware 
implementation of RCP, \cite{piqi} develops a new congestion controller called Proportional Integral Queue Independent RCP,  \cite{gra} performs local bifurcation analysis for some congestion control algorithms, and \cite{zhang} develops an experimental framework used to evaluate several ECPs.  The range of these studies exhibit the difficulty 
of developing a new transport protocol.

RCP routers obtain rate estimates from two forms of feedback: one is based on rate 
mismatch and the other is from the queue size. Thus far the role of both forms of feedback has 
not been well understood. \cite{nanswitch} provides some understanding of the role of queuing dynamics
on the local stability of explicit congestion control algorithms. Nevertheless, there are still no
clear guidelines on choosing vital protocol parameters that influence 
 stability and link utilization.

In this paper, we study two nonlinear fluid models of RCP and analyze the role played by queue feedback
in RCP performance. We claim, using packet level simulations from one of our collaborators in \cite{krv}, that non-switched system is more 
appropriate and in this regime we develop necessary and sufficient conditions for local stability. 
We also computationally study the inherent nonlinear aspects of the protocol.
We then study another model for RCP, where the queue is not modeled as 
a separate fluid quantity, but is a deterministic representation for the underlying stochastics. In this model, 
we further develop the understanding of RCP by considering local stability and a local bifurcation analysis. 
This is done by varying the network parameters, which impact the stability and the link utilization. 

The rest of the paper is structured as follows. In Section II, we outline and analyze two models 
for RCP. In Section III, we summarize our contributions and conclude the paper.

\section{Stability Analysis of two RCP models}

In this section, we outline two nonlinear dynamical systems models for the RCP \cite{nanswitch, krv}. These models have previously been motivated with the objective to 
help design and better understand the performance of RCP.

In the operation of RCP, the feedback from the routers to the end-systems is time-delayed which makes 
it important to understand the stability properties of the nonlinear models. 
Here, we develop both necessary and sufficient conditions for stability of RCP that are not previously shown in literature, and explore the 
consequences of such local stability conditions being violated, both analytically and numerically,
where bifurcation phenomena may readily occur.

\subsection{Model A}

The protocol strives to estimate the fair rate through a single bottleneck link from the rate mismatch 
and the queue size. In order to understand the performance of the protocol, the following nonlinear 
dynamical system for the rate and the queue has been proposed~\cite{nanswitch, FCT}:

\begin{equation}
\frac{d}{dt}\, R(t)= \frac{ R(t)}{C \overline{T}}
\left( a (C - y(t)) - \beta \frac{ q(t)}{ \overline{T}}
\right)
\label{eq:rate}
\end{equation}
where
\begin{equation}
y(t) = \sum_{s} R(t-T_{s})
\label{eq:y}
\end{equation}
and
\begin{equation}
\label{eq:queue}
\begin{aligned}
\frac{d}{dt}q\left(t\right) & = \left[y\left(t\right) - C\right] &\qquad q\left(t\right) > 0\,\\
&= \left[y\left(t\right) - C\right]^{+} &\qquad q\left(t\right) = 0,
\end{aligned}
\end{equation}
using the notation $x^{+} = \max(0, x)$.
Here $R(t)$ is the rate being updated by the router,
$C$ is the link capacity,
$y(t)$ is aggregate load at the link,
$q(t)$ is the queue size,
$T_s$ is the round trip time (RTT) of flow $s$,
and $\overline{T}$ is the average round trip time,
over the flows present. In the formulation of the RCP equation (\ref{eq:rate}), $a$ and $\beta$ are non-negative dimensionless parameters. It is important to understand the impact that these parameters would have on the performance of the protocol.  

The nonlinear rate equation~(\ref{eq:rate}) utilizes two
forms of feedback: one for the rate mismatch which is characterized by
$C - y\left(t\right)$, and another for the instantaneous queue size, $q\left(t\right)$. 
The rate mismatch term causes the rate to increase if the utilization is lower than the 
link capacity $C$ and the queue feedback term serves to decrease the feedback rate as the 
queue size in the router starts to build up.     

Some sufficient conditions for the local stability of the
system~(\ref{eq:rate}), (\ref{eq:y}), (\ref{eq:queue}),
about its equilibrium point, were derived in~\cite{nanswitch},
using techniques developed for a ``switched'' linear control system
with a time delay. The analysis takes into account the discontinuity in the 
system dynamics which would occur as the queue size approaches zero. 
However, we note that this analysis applies to the fluid model rather than a packet level description of the protocol.  

The sufficient conditions, on the non-negative
and dimensionless parameters $a$ and $\beta$, take the functional form
\begin{equation}
a < \frac{\pi}{2}
\label{krveq:cond0a}
\end{equation}
and $\beta < f(a)$ where $f(\cdot)$ is a positive function
that depends on $\overline{T}$.        

We refer to some packet level simulations performed for RCP using a discrete event simulator,
reported in \cite[Fig.~5]{krv}. These simulations demonstrate that the queue, in equilibrium, may not sit at zero but rather the mean queue size would be close to it. The authors consider a network consisting of a single bottleneck link with a capacity $C$ of one packet per unit 
time and 100 Poisson sources and RTT 100 time units. The parameters with queue 
feedback are: $a = 0.5$, $\beta =1$ for a utilization of 90\%; the parameters without queue feedback 
are: $a = 1$, $\beta = 0$, with capacity $\gamma C$, where $\gamma = 0.9$ to target a utilization of 90\% as before. It was observed that with queue feedback the queue size is unstable and has oscillations. On the other hand, 
without queue feedback, the queue size term is 
stable and has a non zero equilibrium value. 

This motivates our subsequent 
stability analysis for necessary and sufficient conditions.
Clearly, the two forms of feedback are playing a non-trivial 
role which would not be apparent from a linear system. 
Due to this, we also need to understand the protocol behavior when conditions for stability are violated.

\subsubsection{Stability analysis}
We shall now understand the local stability of Model A. Since we wish to focus on the nonlinearity
of the model, we consider the following modified equation for the queue dynamics, instead of (\ref{eq:queue})

\begin{equation}
\label{eq:modq}
\begin{aligned}
\dot{q}\left(t\right) & = \left[y\left(t\right) - C\right] \qquad \forall q\left(t\right).
\end{aligned}
\end{equation}

Assume that all flows through the bottleneck link have a common RTT, $T$. Our system 
is now represented by (\ref{eq:rate}), (\ref{eq:y}), (\ref{eq:modq}). The fixed points for these equations are: 
\begin{equation}
\label{eq:fixedpts}
R^{\star} = C/n,\qquad q^{\star} = 0,
\end{equation}
where $n$ is the number of flows. Upon linearizing (\ref{eq:rate}), (\ref{eq:y}), 
(\ref{eq:modq}) about the fixed point and on further simplification, we get
\begin{equation}
\begin{aligned}
\label{eq:linear}
\dot{r}\left(t\right) &= -\frac{a}{T}\left(r\left(t-T\right)\right)-\frac{\beta}{nT^{2}}q\left(t\right)\\
\dot{q}\left(t\right) &= nr\left(t-T\right),
\end{aligned}
\end{equation}
with $r\left(t\right) = R\left(t\right) - C/n$. Using a Laplace transform, we get
\begin{equation}
\label{eq:charwithBeta}
\begin{aligned}
S^{2}T^{2}e^{TS} + aTS + \beta = 0,
\end{aligned}
\end{equation}
where $S$ is the complex argument in the frequency domain. We consider two cases: when $\beta =0$ and $\beta > 0$ corresponding to queue feedback being absent and present, respectively.

\paragraph{Without queue feedback}
Here, (\ref{eq:charwithBeta}) becomes
\begin{equation}
\label{eq:charwithoutBeta}
\begin{aligned}
STe^{TS} + a = 0.
\end{aligned}
\end{equation}

We let $TS = \lambda$, and introduce a new parameter, $\eta$, in (\ref{eq:charwithoutBeta}):
\begin{equation}
\label{eq:eta1}
\begin{aligned}
\lambda e^{\lambda\eta} + a = 0.
\end{aligned}
\end{equation}

Substituting $\lambda = j\omega$, and comparing real and imaginary parts above, we get $\omega\eta = \left(2m + 1\right)\pi/2$ and $\omega = a$. Hence, $a = \pi/\left(2\eta\right)$ when $m = 1$. We also note that when $\eta= 0$, the only root of 
(\ref{eq:eta1}) is $\lambda = -a$ that is stable. Now, using Rouch\'{e}'s theorem 
\cite{lang}, we find that the system represented by 
(\ref{eq:eta1}) is stable if $\eta < \pi/\left(2a\right)$. 
With $\eta = 1$, we get back our original characteristic equation, and the stability condition becomes  
\begin{equation}
\label{eq:stablenob}
\begin{aligned}
a < \frac{\pi}{2}.
\end{aligned}
\end{equation}

This is a 
necessary and sufficient condition for local stability for the system  
(\ref{eq:rate}), (\ref{eq:y}), (\ref{eq:modq}) when $\beta = 0$, which implies that there 
is no queue feedback.

\paragraph{With queue feedback}
Here, (\ref{eq:charwithBeta}) becomes ($TS=\lambda$):

\begin{equation}
\label{eq:characteristic}
\begin{aligned}
\lambda^{2}e^{\lambda} + a\lambda + \beta = 0.
\end{aligned}
\end{equation} 

Once again, we introduce a new parameter $\eta$ as follows:
\begin{equation}
\begin{aligned}
\label{eq:eta} 
\lambda^{2}e^{\eta\lambda} + a\lambda + \beta = 0.
\end{aligned}
\end{equation}

By substituting $\lambda = j\omega$ in the above equation we get:
\begin{equation}
\begin{aligned}
-\omega^{2}\cos\left(\eta\omega\right) + \beta &= 0,\text{ and } 
-\omega^{2}\sin\left(\eta\omega\right) + a\omega &= 0.
\end{aligned}
\end{equation}

Upon simplifying the above equations, we get:
\begin{equation}
\label{eq:omega}
\omega = \pm \sqrt{\frac{a^{2} + \sqrt{a^{4} + 4\beta^{2}}}{2}},
\end{equation}
as $\omega^{2}$ is non-negative. $\omega\sin\left(\eta\omega\right) = a $ as 
$\omega \neq 0$ if $a > 0$, so
\begin{equation}
\begin{aligned}
\eta = \frac{1}{\omega}\sin^{-1}\left(\frac{a}{\omega}\right).
\end{aligned}
\end{equation}

Now, if $a$ and $\beta$ are fixed and $\eta = 0$, the roots of (\ref{eq:eta}) have negative real parts, as $a$ is positive, and are stable. Once again, using Rouch\'{e}'s theorem, and with $\eta = 1$, we get
\begin{equation}
\begin{aligned}
1 < \frac{1}{\omega}\sin^{-1}\left(\frac{a}{\omega}\right)
\end{aligned}
\end{equation}
as the stability condition for the linearized system corresponding 
to equation (\ref{eq:characteristic}). This can be simplified as
\begin{equation}
\begin{aligned}
\tan\frac{\sqrt{a^{2} + \sqrt{a^{4}+4\beta^{2}}}}{\sqrt{2}} < \frac{a}{\beta}\frac{\sqrt{a^{2} + \sqrt{a^{4}+4\beta^{2}}}}{\sqrt{2}},
\end{aligned}
\end{equation}
which is the necessary and sufficient condition for the local stability for the system  
(\ref{eq:rate}), (\ref{eq:y}), (\ref{eq:modq}) when $\beta > 0$; see Fig. 1.
\begin{figure}[t!]
\label{stableA}
\centering
\psfrag{B}{{\small $\beta$}}
\psfrag{A}{{\small $a$}}
\psfrag{Stable}{\footnotesize Stable region}
\includegraphics[width=2.5in]{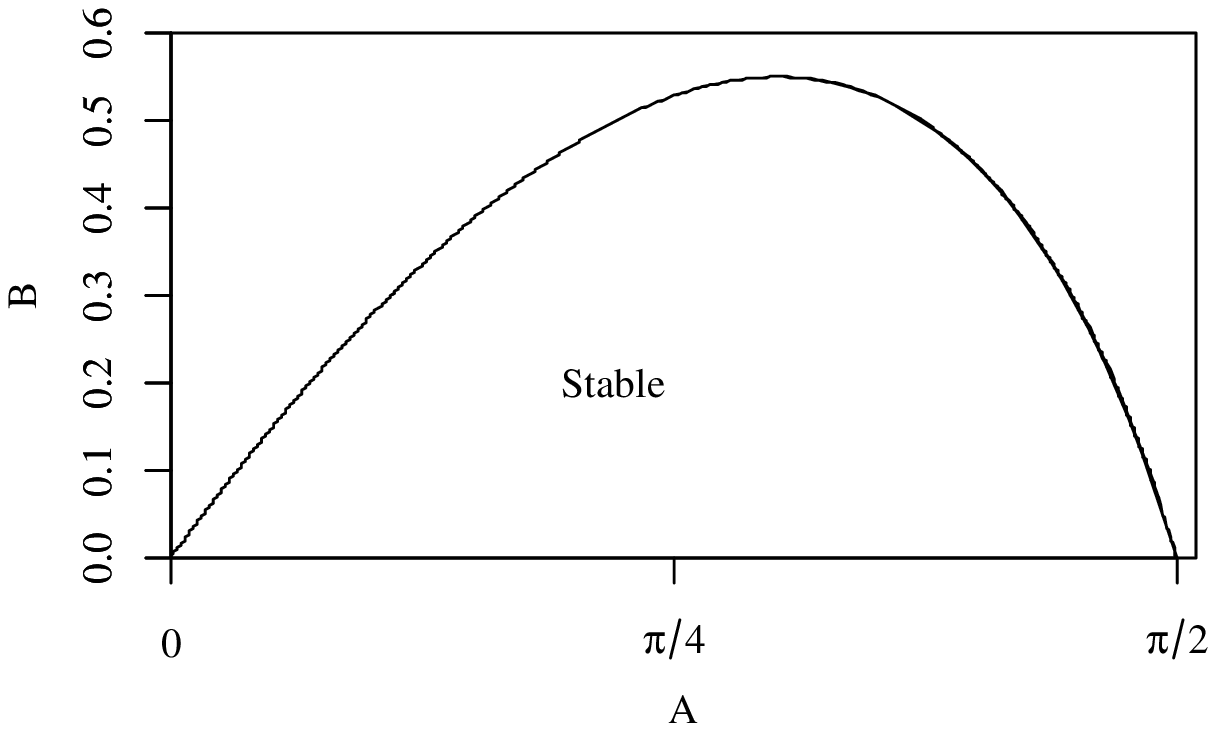}
\caption{Stability chart for Model A}
\end{figure} 

\begin{figure*}
\centering
\psfrag{Ratemmmm}{\small rate, $R\left(t\right)$}
\psfrag{a}{\small $a$}
\psfrag{queuemmmm}{\small queue, $q\left(t\right)$}

\psfrag{a1}{\small $a = 1.9$}
\psfrag{a2}{\small $a = 1.5$}
\psfrag{a3}{\hspace{0.1cm} \small $a = 1.1$}

\psfrag{b1}{\small $a = 3.3$}
\psfrag{b2}{\small $a = 1.8$}
\psfrag{b3}{\hspace{0.11cm} \small $a = 1$}

\includegraphics[width=7in]{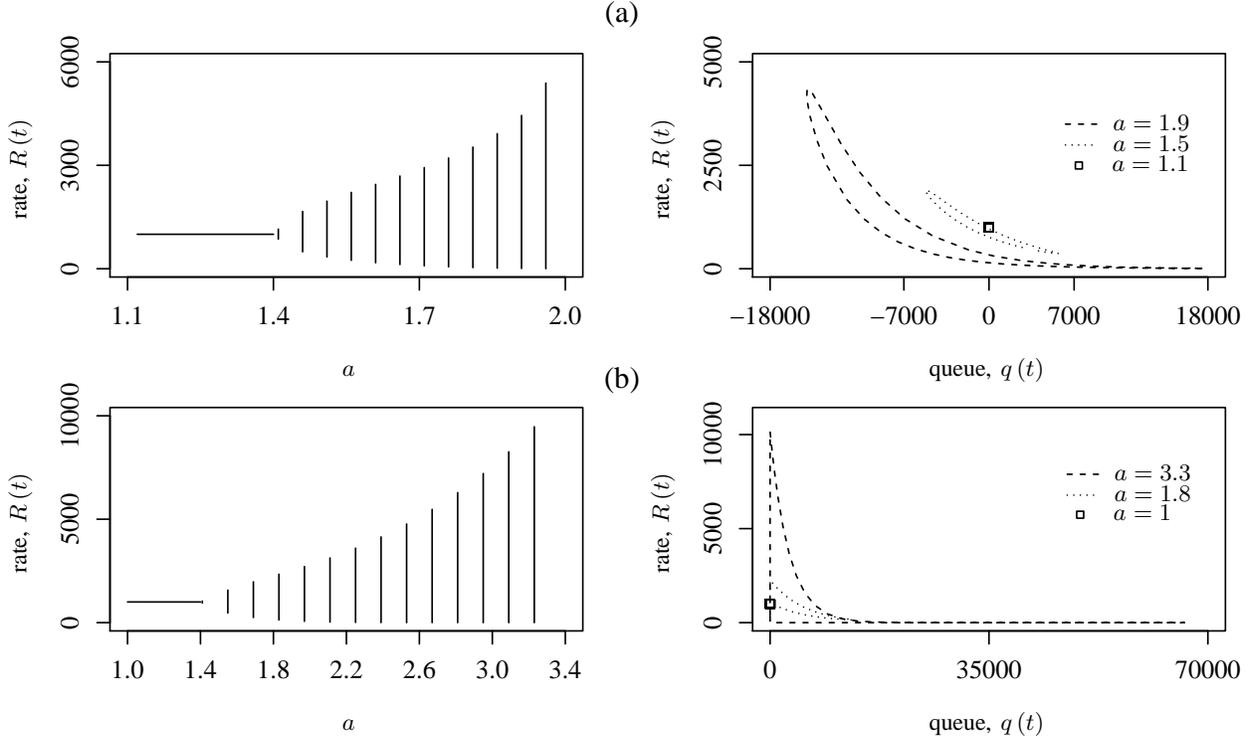}
\caption{Numerical computations for Model A: (a) Bifurcation diagram (left) and phase portrait 
(right) for the ``non-switched'' case using (\ref{eq:modq}), with $\beta = 0.3$ and $a$ is varied, (b) Bifurcation diagram 
(left) and phase portrait (right) for the ``switched'' case using (\ref{eq:queue}), with $\beta = 0.3$ and $a$ is 
varied. Other parameter values are: RTT, $T = 0.1$ time units; Capacity, $C = 100,000$ packets per 
unit time; Number of flows, $s = 100$.}
\label{multiplot}
\end{figure*}

\subsubsection{Numerical results}
We performed numerical computations for Model A and presented the results in Fig. 2.
We chose $\beta = 0.3$, RTT $T = 0.1$ time units,
capacity $C = 100,000$ packets per unit time, number of flows $s = 100$, and varied the parameter $a$. We plotted the bifurcation
diagrams and the phase portraits, both with and without the switch in the queue dynamics. The results confirm our analytically derived stability region, with the system becoming unstable when $a$ leaves the stability region. Additionally, the results show the nonlinear nature of the instability in the form of limit cycles, which are visually apparent in the phase portraits. The cycle for $a = 1.1$, in the non-switched case, and
$a = 1$, in the switched case, are so small that they had to be represented using a square box.

Note that the dynamics of Model A undergo a bifurcation at the boundary of the stability region.
The analysis in Section-IIA1 is done for the system (\ref{eq:rate}), (\ref{eq:y}), (\ref{eq:modq}).
We have neglected the switching in the dynamics of the queue size, using (\ref{eq:modq}), to highlight the nonlinearity in the
dynamics of the rate term. Nevertheless, the actual behavior of RCP is better
represented by the system (\ref{eq:rate}), (\ref{eq:y}), (\ref{eq:modq}). \cite[Fig.~5]{krv}
shows that the mean queue size is not zero.    
Hence, the switch in the queue size dynamics does not play a vital role, since the equilibrium queue size is non-zero.
Furthermore, we observe from Fig. 2 that the dynamics of the rate term undergoes a bifurcation with respect to the parameter $a$, both with and without switching in the queue dynamics. The phase
portraits show that both cases are topologically equivalent as we can
obtain one from the other by distorting the phase portrait appropriately. It is,
therefore, reasonable to analyze Model A without the switch in the queue size dynamics.

We now present another model used to represent RCP.

\subsection{Model B}
Model A accounts for the queue term explicitly via the differential equations (\ref{eq:queue}), (\ref{eq:modq}).
We now outline a small buffer model of RCP. In this regime, the queue size fluctuates so rapidly that it becomes impossible to respond to and 
control its actual size. Instead, RCP behaves as if it is responding to a \emph{ 
distribution} of the queue size. Therefore, at the time scale pertinent for the convergence of
the system, the \textit{mean} queue size is more important. It is also assumed that the queuing delay is negligible 
compared to the propagation delay, which conforms with the small buffer assumption.

A small queue variant of RCP that is proportionally fair is described by the following 
nonlinear differential equations \cite{krv}   

\begin{equation}
\label{eq:ratemodelB}
\frac{d}{dt}R_{j}\left(t\right) = \frac{aR_{j}\left(t\right)}{C_{j}\overline{T}_{j}\left(t\right)}\left(C_{j} - y_{j}\left(t\right) - b_{j}C_{j}p_{j}\left(y_{j}\left(t\right)\right)\right)
\end{equation}
where 
\begin{equation}
\label{eq:modelB_2}
y_{j}\left(t\right) = \sum_{r:j\in r} x_{r}\left(t - T_{rj}\right)
\end{equation}
is the aggregate load at resource $j$ summed over all the routes, $r$, containing resource $j$; 
$x_{r}\left(t\right)$ is the flow rate leaving the source of route $r$; $p_{j}\left(y_{j}\right)$ 
is the mean queue size at link $j$ when the load there is $y_{j}$; capacity of resource $j$ is $C_{j}$; and
\begin{equation}
\overline{T}_{j}\left(t\right) = \frac{\sum_{r:j\in r}x_{r}\left(t\right)T_{r}}{\sum_{r:j\in r}x_{r}\left(t\right)}
\end{equation}
is the average RTT of packets passing through resource $j$. Here, 
$T_{r}$ is the sum of the propagation delay from the source of the flow on route $r$ to resource $j$ ($T_{rj}$) 
and the propagation delay from the resource $j$ to the source of the flow on route $r$ ($T_{jr}$). In (\ref{eq:ratemodelB}), $a$ and $b_{j}$ are non-negative dimensionless parameters. Let the flow rate
 $x_{r}\left(t\right)$ leaving the source of route $r$ at time $t$ be given by

\begin{equation}
x_{r}\left(t\right) = w_{r}\left(\sum_{j\in r}R_{j}\left(t - T_{jr}\right)^{-1}\right)^{-1},
\end{equation}
where $w_{r}$ is the weight given to route $r$.
We can obtain an expression for the mean queue size in the following way: consider the 
arriving workload at resource $j$ is Gaussian over a time period $\tau$, with mean 
$y_{j}\tau$ and variance $y_{j}\tau\sigma_{j}^{2}$. Then the workload present at the queue 
is a reflected Brownian motion, with mean under its stationary distribution of

\begin{equation}
\label{p}
p_{j}\left(y_{j}\right) = \frac{y_{j}\sigma_{j}^{2}}{2\left(C_{j} - y_{j}\right)}.
\end{equation}

In essence, 
the queue size term is being modeled by $p_{j}\left(y_{j}\right)$ as described 
in (\ref{p}). The parameter $\sigma_{j}$ determines how variable the 
traffic at resource $j$ is. For instance, if $\sigma_{j} = 1$, then the traffic is Poisson. 

We should note that the parameter $a$ in both the models is the same. The parameter $b_{j}$ of 
Model B can be related to the parameters $a$ and $\beta$ of Model A by the equation

\begin{equation}
\label{eq:b}
b_{j} = \frac{\beta}{aC_{j}\overline{T}_{j}}.
\end{equation}

The parameter $b_{j}$ affects the utilization of resource $j$ at equilibrium. From (\ref{p}) 
and considering the condition for equilibrium of the system in (\ref{eq:ratemodelB}), (\ref{eq:modelB_2}), we can evaluate the utilization of resource $j$, $\rho_{j}$, after simplification, as
\begin{equation}
\label{eq:util}
\rho_{j} = 1 - \sigma_{j}\sqrt{\left(\frac{b_{j}}{2}\right)} + O\left(\sigma_{j}^{2}b_{j}\right).
\end{equation}

\subsubsection{Stability analysis}
In \cite{krv}, a sufficient condition for the local stability of Model B for heterogeneous 
propagation delays was derived. It was shown that with queue feedback, a sufficient condition for stability is $a < \pi/2$, while without queue feedback, it is $a < \pi/4$.

Here, we present the necessary and sufficient conditions for the local stability of this model, for 
a homogeneous propagation delay, both with and without queue feedback. We note 
that we get similar sufficient conditions as mentioned in \cite{krv}. We also prove, analytically, 
the existence of a Hopf bifurcation at the edge of the stability region in both cases.

\paragraph{With queue feedback}

Let the network 
consist of a single link with capacity $C$, a single route and a common RTT, $\tau$, for all the flows. We drop the subscripts, $j$ and $r$, for clarity. 
All the flows send Poisson traffic, hence, $\sigma = 1$. We take $w = 1$ as this 
only affects the equilibrium point. 
For this scenario, the general rate equation in (\ref{eq:ratemodelB}) becomes

\begin{equation}
\label{eq:simple_rate_B}
\frac{d}{dt}R\left(t\right) = \frac{aR\left(t\right)}{C\tau}\left(C - y\left(t\right) - bCp\left(y\left(t\right)\right)\right)
\end{equation}
where
\begin{equation}
\label{eq:simple_y_B}
y\left(t\right) = R\left(t-\tau\right),\text{ and } p\left(y\right) = \frac{y}{2\left(C - y\right)}.
\end{equation}

We now linearize (\ref{eq:simple_rate_B}) using the Taylor expansion about the 
equilibrium. Let $R\left(t\right) = r\left(t\right) + R^{\star}$, where $R^{\star}$ is 
the equilibrium value of $R\left(t\right)$, and $r\left(t\right)$ is a small perturbation about 
the equilibrium. Hence, (\ref{eq:simple_rate_B}) becomes

\begin{equation}
\label{eq:LinB}
\frac{d}{dt}R\left(t\right) = \dot{r}\left(t\right) = 
-a\left(\frac{R^{\star} + C}{C\tau}\right)r\left(t-\tau\right).
\end{equation}
where $R^{\star} = C\left(b+4-\sqrt{b^{2} + 8b}\right)/4$. More simply,
\begin{equation}
\label{eq:finLinB}
\dot{r}\left(t\right) = -\kappa r\left(t-\tau\right),
\end{equation}
where $\kappa = a\left(2 + b/4 - \sqrt{b^{2}/16 + b/2}\right)/\tau$. By taking the 
Laplace transform of (\ref{eq:finLinB}), we get the characteristic equation

\begin{equation}
\label{eq:charB}
\lambda + \kappa e^{-\lambda\tau} = 0,
\end{equation}
where $\lambda$ is the complex argument in the frequency domain. Once again, we introduce
a new parameter $\eta$ as,
\begin{equation}
\label{eq:charB_eta}
\lambda + \kappa e^{-\lambda\tau\eta} = 0.
\end{equation}

By substituting $\lambda = j\omega$ and comparing real and imaginary parts, we get $\kappa\tau\eta = \pi/2$. We note that when $\eta = 0$, the only 
solution of (\ref{eq:charB_eta}) is $\lambda = -\kappa$, which is stable. Hence, using 
Rouch\'{e}'s theorem, and substituting $\eta = 1$ to get the characteristic equation (\ref{eq:charB}), we note that $\kappa\tau < \pi/2$ is the necessary and 
sufficient condition for the local stability of Model B, which is expanded as
\begin{equation}
\label{eq:stableBns}
a\left(2 + \frac{b}{4} - \sqrt{\frac{b^{2}}{16} + \frac{b}{2}}\right) < \frac{\pi}{2}.
\end{equation}

If we substitute $b \rightarrow 0$ in (\ref{eq:stableBns}), we get 
\begin{equation}
\label{eq:stableBs}
a < \pi/4.
\end{equation}

This is the sufficient condition for local stability of Model B, with queue feedback as 
$b > 0$. This can be seen from Fig. 3, where $\forall b > 0$, the region $a < \pi/4$ falls in the stable region. 

In (\ref{eq:stableBns}), we note that as $b \rightarrow \infty$, $a \rightarrow \pi/2$. This 
can be shown by the fact that 
\begin{equation}
\lim_{b \rightarrow \infty}\left(1 + \frac{b}{4} - \sqrt{\frac{b^{2}}{16} + \frac{b}{2}}\right) = 0.
\end{equation}

\paragraph{Without queue feedback}
Consider the same scenario as in the previous case and choose $b = 0$ in (\ref{eq:simple_rate_B}).
We linearize (\ref{eq:simple_rate_B}), with $b = 0$, as
\begin{equation}
\label{eq:LinBnoqueue}
\dot{r}\left(t\right) = 
-\left(\frac{aR^{\star}}{C\tau}\right)r\left(t-\tau\right).
\end{equation}
where $R^{\star} = C$, in this case. Substituting $\zeta = a/\tau$ in (\ref{eq:LinBnoqueue}), we get the characteristic equation as
\begin{equation}
\label{eq:char_mod_B_noq}
\lambda + \zeta e^{-\lambda\tau} = 0.
\end{equation}

Following the same approach as in Section-IIB1a, we get
\begin{equation}
\label{eq:stableBb0}
a < \frac{\pi}{2}
\end{equation}
as the necessary and sufficient condition for the local stability of the Model B when $b = 0$, 
that is without queue feedback. 

We plot the stability chart for this model in Fig. 3. Note that the region above the curve represents
the stability region for $b > 0$, whereas when $b = 0$, the system is stable iff $a < \pi/2$; hence there is a discontinuity at $b = 0$, as confirmed by (\ref{eq:stableBns}) and (\ref{eq:stableBb0}).

\begin{figure}
\label{stableB}
\centering
\psfrag{b}{{\small $b$}}
\psfrag{a}{{\small $a$}}
\psfrag{hopf}{\hspace{-0.57cm}\footnotesize Hopf bifurcation boundary $\rightarrow$}
\psfrag{Stable}{\footnotesize Stable region}
\psfrag{Unstable}{ \footnotesize Unstable region}
\includegraphics[width=3.5in]{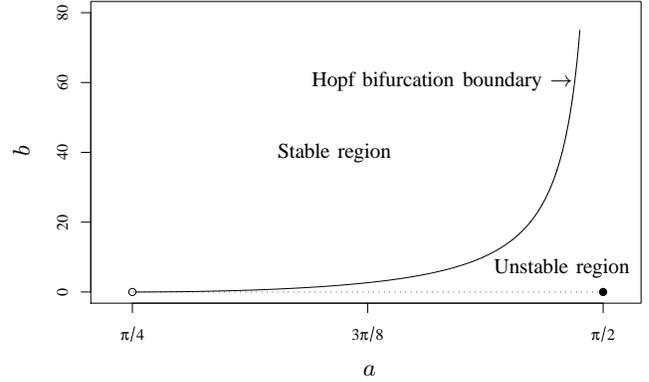}
\caption{ Stability chart for Model B}
\end{figure}

\paragraph{Hopf bifurcation}

We show the system in (\ref{eq:charB}), i.e. with queue feedback, undergoes a Hopf bifurcation as $a$ is varied. Consider (\ref{eq:charB}) by rewriting $\kappa$ as $\kappa = \xi a/\tau$ with $\xi = 2 + b/4 - \sqrt{b^{2}/16 + b/2}$. Then, we get
\begin{equation}
\label{eq:charA}
\lambda +\frac{a\xi}{\tau} e^{-\lambda\tau} = 0.
\end{equation}

Differentiating the above equation with respect to $a$, we get
\begin{equation}
\frac{d\lambda}{da} = \frac{1}{a\tau - \frac{\tau}{\xi} e^{\lambda\tau}}.
\end{equation}

Define $\text{Re}\left(z\right)$ as the real part of $z$, and $\text{sgn}\left(\cdot\right)$ as the sign function. Then,

\begin{multline}
\label{eq:signs}
\text{sgn}\left(\text{Re}\left(\frac{d\lambda}{da}\right)\right) = \text{sgn}\left(\text{Re}\left(\frac{1}{a\tau - \frac{\tau}{\xi} e^{\lambda\tau}}\right)\right) \\
 = \text{sgn}\left(\text{Re}\left(\xi a - e^{\lambda\tau}\right)\right)
 = \text{sgn}\left(\text{Re}\left(a + \frac{a}{\lambda\tau}\right)\right)
 = \text{sgn}\left(a\right) > 0,
\end{multline}
where the second to last equality is obtained by substituting for $e^{\lambda\tau}$ from 
(\ref{eq:charA}), and the last equality is because $\text{Re}\left(\lambda\right) = 0$ at the boundary. This shows that at the boundary of the 
stability region $\text{Re}\left(\frac{d\lambda}{da}\right) \neq 0$, while $\text{Re}\left(\lambda\right) < 0$ inside the region. Hence, $\lambda$ changes signs across the region and the fixed point becomes unstable, i.e. we say that it undergoes a Hopf bifurcation.
The condition $\text{Re}\left(\frac{d\lambda}{da}\right) \neq 0$ is
called the transversality condition for a Hopf bifurcation.

Similarly, the above analysis also applies for the case of no queue feedback with (\ref{eq:char_mod_B_noq}), along the same lines as (\ref{eq:charB}).

\begin{figure*}
\centering
\psfrag{Ratemmmmmm}{{\small rate, $R\left(t\right)$}}
\psfrag{a}{{\small $a$}}
\psfrag{rate,,R(t-T)}{{\small rate, $R\left(t-T\right)$}}

\psfrag{a1}{{\footnotesize $a=0.652$}}
\psfrag{a2}{{\footnotesize $a=0.638$}}
\psfrag{a3}{{\footnotesize $a=0.586$}}

\psfrag{b1}{{\footnotesize $a=0.876$}}
\psfrag{b2}{{\footnotesize $a=0.858$}}
\psfrag{b3}{{\footnotesize $a=0.788$}}

\psfrag{c1}{{\footnotesize $a=1$}}
\psfrag{c2}{{\footnotesize $a=1.4$}}
\psfrag{c3}{{\footnotesize $a=1.6$}}
\psfrag{c4}{{\footnotesize $a=2$}}

\psfrag{d1}{{\footnotesize $a=1$}}
\psfrag{d2}{{\footnotesize $a=1.4$}}
\psfrag{d3}{{\footnotesize $a=1.6$}}
\psfrag{d4}{{\footnotesize $a=2$}}

\includegraphics[height=8in]{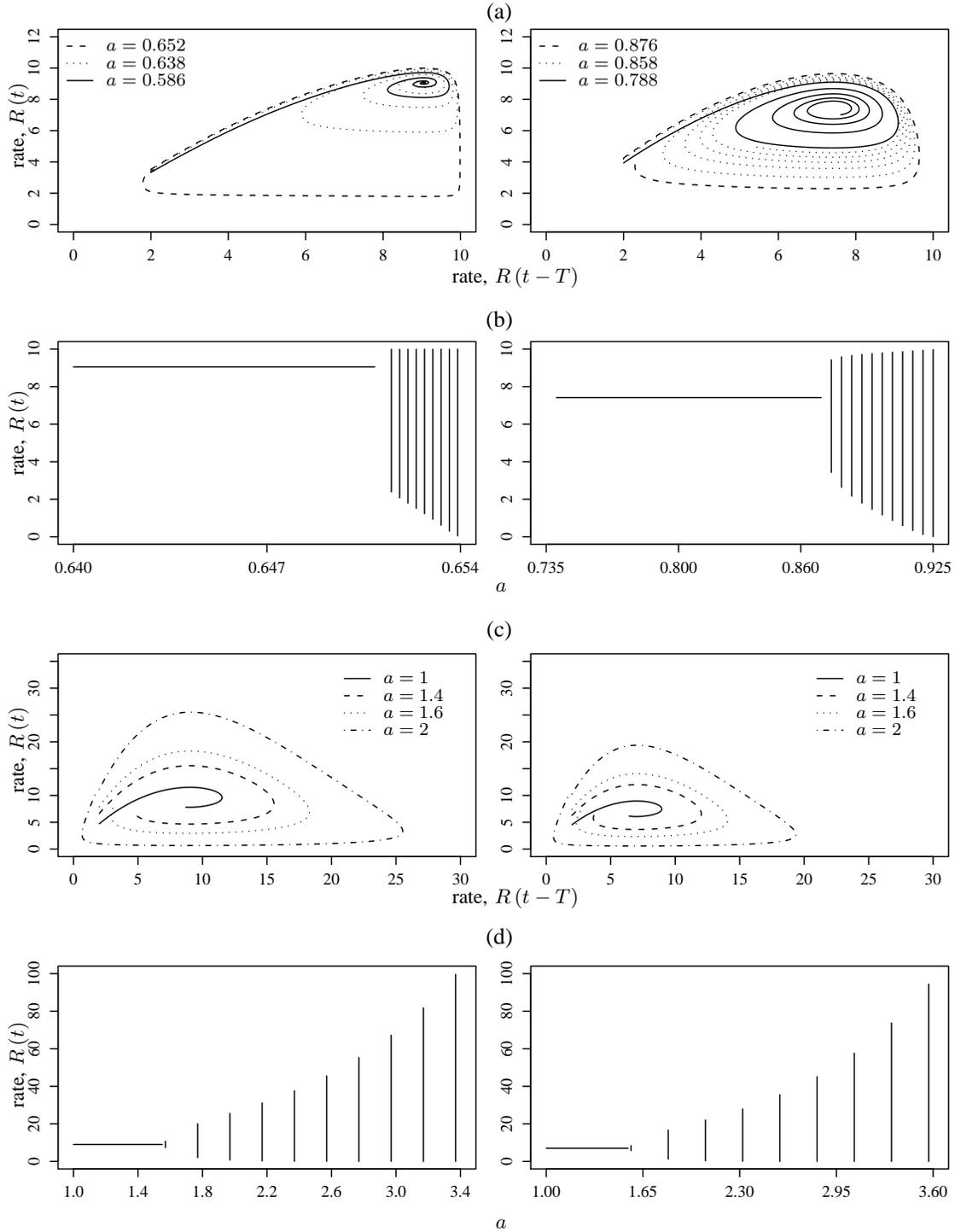}
\caption{Numerical computations for Model B -- (a) Phase portrait, with queue: Utilization = 90\% (left), 
70\% (right), (b) Bifurcation diagram, with queue: Utilization = 90\% (left), 70\% (right), 
(c) Phase portrait, without queue: Utilization = 90\% (left), 70\% (right), (d) Bifurcation diagram, 
without queue: Utilization = 90\% (left), 70\% (right). The values of the parameters are: RTT, $T = 1$ time unit; 
Capacity, $C = 10$ packets per unit time; with queue feedback -- $b = 0.02$ for 90\% utilization , $b = 0.18$ for 
70\% utilization; without queue feedback -- $b=0$, $\gamma = 0.9$ for 90\% utilization, $\gamma = 0.7$ for
 70\% utilization. The values of parameter $a$ in (a) are chosen to be proportionally spaced from the bifurcation boundary.}
\label{4x2_plot}
\end{figure*}

\subsection{Impact of queue feedback}

In context to the parameters chosen for packet level simulation results in \cite[Fig.~5]{krv}, the stability chart in Fig. 1 shows that $a=0.5$ and $\beta = 1$ is outside the provably 
stable region, corroborating the results in \cite[Fig.~5]{krv}. In the absence of queue feedback, to target a utilization of 90\%, $a = 1$, 
$\beta = 0$ and $\gamma = 0.9$ where $C$ is replaced with $\gamma C$. In this case, the RCP model only 
reacts to rate mismatch. We note that the simulations in \cite{krv} did not produce any deterministic instabilities, and our analysis suggests that at these parameters the protocol is locally stable. Therefore, in this regime, the presence of queue feedback causes the queue 
to be \textit{less} accurately controlled, suggesting a fundamental
difference between these two forms of feedback and also showing evidence in favor of no queue feedback.

For model B, we have analytically shown that a Hopf bifurcation arises at the boundary of the stability region, 
which signifies the emergence of limit cycles. Of course, it is important to determine the 
stability of the bifurcating periodic orbit. An analytical characterisation of the stability or instability 
of the bifurcating limit cycle is beyond the scope of this paper. However, the computations performed for 
Model B, shown in Fig. 4, suggest that the limit cycles could indeed be stable. In Fig. 4, 
with queue and without queue feedback, we observed stable limit cycles that were plotted in the corresponding 
bifurcation diagrams. We observe that the corresponding phase portraits 
are topologically equivalent. We would like to highlight that the computations done for Model A, represented in Fig. 2, also provide evidence for the emergence of stable limit cycles as protocol parameters are varied. 
To that end, the packet level simulations shown in \cite{krv}, which exhibit nonlinear oscillations appear to be 
induced via a Hopf type bifurcation. Our computational and analytical results for the Hopf 
bifurcation together provide rather strong evidence for a thorough investigation on the nonlinear dynamical 
characteristics for the various RCP models outlined in this paper. 

\subsection{Impact of utilization}
We shall now look 
at how varying utilization impacts stability. From Fig. 4(b) and 4(d), we see that as utilization 
is decreased, the system enters a limit cycle at a later value of parameter $a$. The amplitude 
of the limit cycle also grows slower. This effect is more pronounced with queue feedback. 
This suggests to us that as utilization is decreased, the stability of the system increases. 

\subsection{Impact of parameter $a$}
We observe in Models A and B that the parameter $a$ appears along with the rate mismatch term, 
$C - y\left(t\right)$. Due to this, parameter $a$ affects the speed at which 
the equilibrium rate is attained as the magnitude of the rate mismatch feedback changes with 
$a$. We now show this using theoretical analysis.

Note that if $\text{Re}\left(\frac{d\lambda}{da}\right) < 0$ and $\text{Re}\left(\lambda\right) < 0$, 
then we observe that the solution of the linearized system in (\ref{eq:finLinB}) decays faster as $a$ increases. This motivates our analysis. Using the second to last expression in (\ref{eq:signs}), the condition for $\text{Re}\left(\frac{d\lambda}{da}\right) < 0$ is
\begin{equation}
\text{Re}\left(\lambda\right) > -\frac{1}{\tau}.
\end{equation}

We solve (\ref{eq:char_mod_B_noq}) for $\lambda = -1/\tau$ and obtain $a = 1/e$. We also note that when 
$a = 0$, $\lambda = 0 > -1/\tau$, hence 
$\text{Re}\left(\frac{d\lambda}{da}\right) < 0$. Obviously, there exists a particular value of $a$ after which 
$\lambda < -1/\tau$ and this value of $a$ is $1/e$. Therefore, we note that the rate of 
convergence increases upto $a = \frac{1}{e}$, and then decreases again beyond that. Simulations confirming this analysis have been presented in 
\cite{NanditaPhD}.

Furthermore, we note again that parameter $a$ also affects the stability of the system, as shown by our analysis in this section and in Fig. 2 and 4, and causes limit cycles for sufficiently large $a$. From the above analysis, we conclude that the optimal value of $a$ for the fastest rate of convergence is
\begin{equation}
a = \frac{1}{e}.
\end{equation}

\section{Conclusions}                  

Our focus, in this paper was on the equilibrium properties of two specific models of RCP. These two nonlinear models have only been studied for sufficient conditions to ensure local stability. To that end, for both these models, we developed necessary and sufficient conditions for local stability, under certain conditions. As conditions for stability get violated, bifurcations may occur. For both the models, we explored the consequences of parameters violating the stability conditions, and we plotted the respective bifurcation plots for the emerging stable limit cycles. For the small buffer variant of RCP, we also analytically showed that the bifurcation would be a Hopf bifurcation, which does signify the emergence of limit cycles. We used this insight, to help explain the potential destabilizing effect of having two forms of feedback in the protocol definition of RCP.    

This, we believe, sheds light on a key architectural question concerning the design of RCP, i.e. whether the protocol needs to estimate the fair rate from both rate mismatch and from the queue size. Hopf type bifurcations, occurring due to the presence of queue size in the feedback, open additional questions regarding the nonlinear properties of the fluid models and their relationship with protocol design.


\section*{Acknowledgment}
This work was performed while the author was affiliated with the Indian Institute of Technology, Madras, India, working under the guidance of Dr. G. Raina. The author expresses gratitude for the helpful discussions pointing to reference \cite{krv}.

\balance
\bibliographystyle{IEEEtran}
\bibliography{IEEEabrv,refs}
\end{document}